\documentclass[aps,prd,twocolumn,groupedaddress]{revtex4}
\usepackage{graphicx}

\begin{document}

\title{ A  Roadmap For Meson Spectroscopy }
\author{M.G. Olsson}
\affiliation{Physics Department, University of Wisconsin, Madison, WI
53706}

\begin{abstract}
An efficient classification of light quark meson states is discussed based
on the dominance of angular and radial quark excitation.  A synthesis of
Regge and quark dynamics allows a natural unification of light
quark-antiquark spectroscopies and indicates the states that fall outside
this category such as molecules, hybrids, and glueballs.
\end{abstract}

\maketitle

\section{Introduction}

The rapidly evolving field of hadron spectroscopy is increasingly motivated
by firm expectations of seeing new types of matter\cite{key}.  The
``ordinary" mesons are thought to be confined quark-antiquark pairs in spin
singlet/triplet states and with orbital and radial excitation.
Non-perturbative QCD in addition predicts that for each ordinary meson
there are higher lying states with excited gluonic energy (hybrids) and
that there should also be mesons composed essentially of pure glue energy
(glueballs).  Further complicating the experimental situation are possible
four quark mesons (molecules) or dynamically generated resonances not
having any clear relation to the quark picture.

In the most recent Particle Data Group report \cite{pdg} there are about
100 well established meson states composed of light quarks $(u,d,s)$ and
glue.  Of these, 56 have been assigned into constituent quark multiplets
\cite{pdg}.  Since the vast majority of the observed meson states lie in 
energy below 2.5~GeV a straightforward method of separating the
ordinary mesons from hybrids, glueballs etc.\ would be very
useful. Previous work \cite{godfrey} in understanding the meson spectrum
has either centered on detailed quark models or on the search for SU(3)
flavor nonets.  Both approaches are sensitive to details such as mixing
between nonets or with hybrids and glueballs.  Although such considerations
are crucial to a complete understanding, the proposed classification is
based on some simple observations and is relatively model independent.

First we draw some motivation from the well-understood heavy onia mesons
composed of $b$ and $c$ quarks.  These states are classified by the orbital
angular momentum and radial excitation of the heavy quarks.  The spin
dependence is much smaller than the quark motion energies and therefore
each state of orbital and radial excitation consists for $L>0$ of four
nearly degenerate states (e.g.\ $\chi_0$, $\chi_1$, $\chi_2$ and $h$).  For
$L=0$ the group is just two states.  The corresponding Regge trajectories
look very different when plotted conventionally as total angular momentum
$J$ vs.\ $E^2$.  Here it is evident that there is a ``fundamental" orbital
angular momentum trajectory $L=L(E^2)$ that breaks into a family of
trajectories when the spin is turned on.  Nothing spectacular happens in
the case of light quarks except that the spin splittings are somewhat
larger.  Thus our first principle will be to consider all light quark
states of the same orbital and radial excitation quantum numbers to form
natural groups.  For $L>0$ and three light quarks there are expected to be
sixteen members of each group (and eight members for $L=0$ groups).
Clearly, a few such groups accounts for a lot of mesons.

For light quarks (and actually also true for heavy quark bound states) the
confining interaction dominates the meson dynamics.  The confining
interaction closest to QCD is the QCD string/flux tube.  The flux tube has
all the proper QCD relativistic corrections \cite{collin} as well as the
correct Regge slope of $1/2\pi a$ \cite{dan}.  A string tension of a=
0.16~GeV$^2$ yields the universal Regge slope of 1~GeV$^{-2}$. In addition
the long range excited glue spectrum is correctly predicted\cite{ted}.  The
light quark QCD flux tube spectroscopy can be exactly numerically
calculated\cite{collin,ted} and analytically accurately estimated
\cite{charlie} by,
\begin{equation}
E^2/2\pi a = L+2n-1/2 \,,
\end{equation}
where $E$ is the meson excitation energy and $L$ takes the values
$0,1,2,$\dots\ and $n=1,2,\dots$\,. We should note the factor $ L+2n$ which
implies degenerate tower states of either even or odd $L$.

The effect of an attractive short-range interaction will be to lower $E$
for a given $L$.  For large $L$ the short-range interaction, $V(r) = -k/r$,
shifts~\cite{coulomb} the Regge trajectory to smaller $E^2/(2\pi a)$ by
$k/2$.  For $L=0$ this shift is somewhat larger.  In Fig.~1 we show by the
solid lines the Regge trajectories due to confinement alone.  These
trajectories given by Eq.~(1) consist of the leading $(n=1)$ and evenly
spaced daughter trajectories all of slope $1/2\pi a$ (or unit slope in this
graph).  This structure is numerically accurate for a straight string and also
for a dynamically curved string~\cite{ted2}.  The dashed curves represent
mesons with both confinement and with short range (taking $k=1$)
interaction.

\begin{figure}
\includegraphics[width=3.5in]{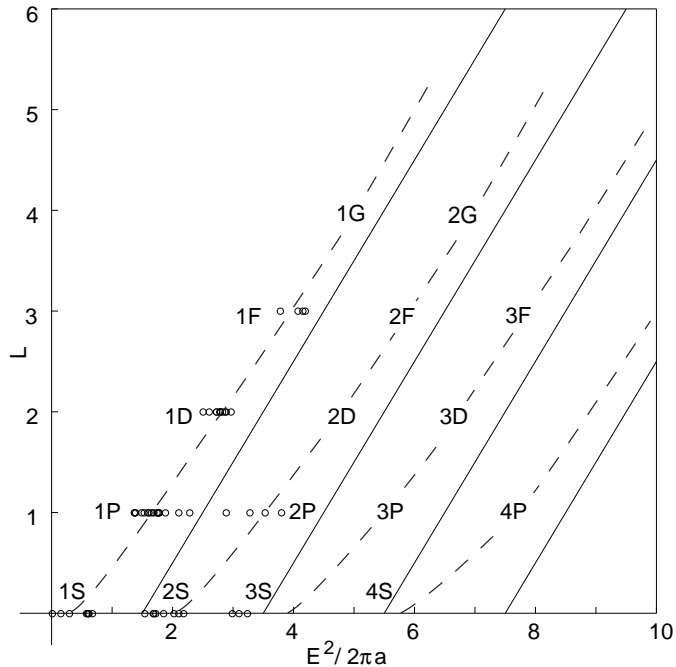}
\caption{ The fundamental Regge plot for light quark mesons.  The ``data"
points represent the 56 quark assignments of the PDG~\cite{pdg}  with
$2\pi a=1$~GeV$^2$.  One hundred MeV has been subtracted for each strange
quark in these mesons. The solid lines are the string confinement
spectroscopy of Eq.(1).  The dashed curves are confinement plus the short
distance Coulomb term $V(r)=-k/r$ with $k=1$~\cite{coulomb}.}
\end{figure}

\section{Ordinary Mesons}

We first consider the light quark mesons \cite{pdg} that have been assigned
to constituent quark configurations.  We subdivide these according to their
assigned orbital and radial quantum numbers $L$ and $n$ \cite{pdg}.  Taking
$2\pi a=1$~GeV$^{-2}$ we plot $E^2/2\pi a$ for each of these mesons in
Fig.1.  In this calculation we have subtracted 0.1~GeV for each strange
quark to roughly account for the differing quark masses.

The ``fundamental" Regge plot of Fig.1 shows that all of the 56 assigned
quark model states indeed fall into a few well defined groups.  All of the
expected states of the 1S, 2S, and 1P groups are present (32 mesons).  The
1D group has currently 12 of the 16 expected and the 3S 4 of 8 states.  We
see that the 2P, and 1F groups each has 4 out of the expected 16.  One
expects the remaining mesons in these groups to be present.  We also
indicate several groups expected at higher energies.

  \section{New  Matter}

In the conservative meson summary table \cite{pdg} there are several mesons
that have no logical place in Fig 1.  The $f_0(.60)$ is probably a
dynamical state and the $f_0(.98)$ and $a_0(.98)$ are surely some sort of
four quark/molecule state associated with the $K\overline K$ threshold.
Much has been written on the nature of these states \cite{pdg}.  Between 1
and 2 GeV there are the 1P, 2S, 1D, 3S, and 2P groups.  Although there are
numerous missing states in these groups there is no place for the
$f_0(1.50)$ which may well be a glueball.  There is also mounting evidence
for two exotic mesons of $J^{PC} = 1^{-+}$ (at 1.4 and 1.6 GeV) which by
their quantum numbers cannot be ordinary mesons.  One of these is an
excellent candidate for a hybrid meson.  Considerable further speculation
is certainly possible but will not be pursued here.

\section{Summary}

Using some general organizing concepts we outline the number, quantum
numbers, and approximate energies of the expected quark-antiquark
(ordinary) mesons.  From orbital and radial motion dominance we find that
all (56) of the mesons currently \cite{pdg} assigned as constituent quark
model states fall into well defined groups as shown in Fig.1.  For massless
(ultra-relativistic) quarks and with linear confinement the allowed
eigenstates generally lie on straight Regge trajectories with parallel
daughter trajectories~\cite{mgo}.  The QCD string seems to be similar to
QCD at large distances giving the simple spectrum (1).  As seen in Fig.~1,
the observed meson groups organize neatly into these fundamental
trajectories when supplemented by the expected short range interaction.
These groupings represent, in a model-independent way, what one might
expect in the way of ordinary mesons.  This should be a useful tool in the
identification of the new matter states.

\acknowledgments 
This research was supported in part by the U.S.~Department
of Energy under Grant No.~DE-FG02-95ER40896 and in part by the University
of Wisconsin Research Committee with funds granted by the Wisconsin Alumni
Research Foundation.


\begin{thebibliography}{99}


\bibitem{key} Key Issues in Hadron Physics, Town Hall Meeting, Jefferson
Laboratory, Dec.\ 1-4 2000, ArXiv:hep-ph/0012238.

\bibitem{pdg} Particle Data Group, K.Hagiwara {\it et al.}, Phys.\ Rev.\
{\bf D66}, 010001 (2002).

\bibitem{godfrey} S. Godfrey and N. Isgur, Phys.\ Rev.\ {\bf D32}, 189
(1985); S. Godfrey and N.~Isgur, Phys.\ Rev.\  {\bf D31}, 2375 (1985);
F.L.~Close, ``Quarks and Nuclear Forces" (Springer-Verlag, 1982).

\bibitem{collin} Collin Olson, M.G.~Olsson, and Ken Williams, Phys.\ Rev.\
{\bf D45}, 4307 (1992); N. Brambilla and G.M. Prosperi, Phys.\ Lett.\ B235,
69 (1990).

\bibitem{dan} D. LaCourse and M.G.~Olsson, Phys.\ Rev.\ {\bf D39}, 2751
(1989); M.G.~Olsson and Sinisa Veseli, Phys.\ Rev.\  {\bf D51}, 3578 (1995);
Yu.\ Dubin, A.B. Kaidalov, and Yu. A. Siminov, Phys.\ Lett.\ {\bf B323}, 41
(1994); {\bf B343}, 310 (1995).

\bibitem{ted} T.J. Allen, M.G. Olsson, and Sinisa Veseli, Phys.\ Lett.\  {\bf
B434}, 110 (1998).

\bibitem{charlie} T.J. Allen, C. Goebel, M.G. Olsson, and Sinisa Veseli,
Phys.\ Rev.\ {\bf D64}, 094011 (2001).

\bibitem{coulomb} 
The Coulomb shift is most simply established for large
$L$ for the classical (leading) Regge trajectory.  For massless quarks and
circular orbits the QCD string~\cite{collin} gives $L=\pi ar^2/8$ and
$E=\pi ar/2 -k/r$.  Eliminating the quark separation $r$, squaring $E$ and
dropping the small $1/L$ term gives $E^2/2\pi a = L-k/2 $.

\bibitem{ted2} T.J.Allen, M.G. Olsson, and Sinisa Veseli, Phys.\ Rev.\ {\bf
D59}, 094011 (1999); Phys.\ Rev.\ {\bf D60}, 074026 (1999).

\bibitem{mgo} M.G. Olsson, Phys.\ Rev.\ {\bf D55}, 5479 (1997).

\end{thebibliography}
\end{document}